\documentclass[preprint,a4paper,10pt,oneside,onecolumn]{elsarticle}

\usepackage{epsfig}
\usepackage{amsmath}
\usepackage{amssymb}
\usepackage{amsfonts}
\usepackage{graphicx}
\usepackage{graphics,subfigure}
\usepackage{float}
\usepackage{booktabs}
\usepackage{color}
\usepackage{geometry}
\usepackage{rotating}

\newcommand{\newc}{\newcommand}
\newc{\ba}{\begin{array}}
\newc{\ea}{\end{array}}
\newc{\bea}{\begin{eqnarray}}
\newc{\eea}{\end{eqnarray}}
\newc{\beastar}{\begin{eqnarray*}}
\newc{\eeastar}{\end{eqnarray*}}
\newc{\beq}{\begin{equation}}
\newc{\eeq}{\end{equation}}
\newc{\bestar}{\begin{equation*}}
\newc{\eestar}{\end{equation*}}
\newc{\ben}{\begin{enumerate}}
\newc{\een}{\end{enumerate}}
\newc{\bi}{\begin{itemize}}
\newc{\ei}{\end{itemize}}

\newc{\lam}{\lambda}
\newc{\lamp}{\lambda^\prime}
\newc{\lampp}{\lambda^{\prime\prime}}
\newc{\Lam}{\Lambda}
\newc{\BLam}{{\mathbf{\Lam}}}
\newc{\eps}{\epsilon}
\newc{\kap}{\kappa}

\newc{\ra}{\rightarrow}
\newc{\ovl}{\overline}
\newc{\lsim}{\stackrel{<}{\sim}}
\newc{\Tr}{{~\rm Tr}}

\newc{\itRahmen}[2]{
\begin{center}\fbox{\parbox{#1 cm}{\it #2}}
\end{center}}

\newc{\del}{\partial}
\newc{\veva}{\langle H_1\rangle}
\newc{\vevb}{\langle H_2\rangle}
\newc{\onehalf}{\textstyle \frac{1}{2} \displaystyle}
\newc{\onethird}{\textstyle \frac{1}{3} \displaystyle}
\newc{\mzero}{M_0}
\newc{\mhalf}{{M_{1/2}}}
\newc{\tanb}{\tan\beta}
\newc{\Psix}{{\mathrm{P}_{\!6}}}
\newc{\nPsix}{{\not\!\Psix}}
\newc{\nPsixU}{{\s{\mathrm P}_{\!6}}}

\newc{\muon}{\mu}
\newc{\azero}{A_0}
\newc{\neutralino}{\tilde\chi^0}
\newc{\selectron}{\tilde e}
\newc{\stau}{\tilde\tau}
\newc{\smuon}{\tilde\mu}
\newc{\sneu}{\tilde\nu}
\newc{\higgs}{h^0}
\newc{\sgnmu}{\textrm{sgn}(\mu)}
\newc{\gev}{\mbox{~GeV}}
\newc{\tev}{\mbox{~TeV}}
\newc{\gsim}{\stackrel{>}{\sim}}
\newc{\mgut}{{M_{GUT}}}
\newc{\mweak}{{M_{W}}}

\newbox\charbox
\newbox\slabox
\def\s#1{{      
    \setbox\charbox=\hbox{$#1$}
    \setbox\slabox=\hbox{$/$}
    \dimen\charbox=\ht\slabox
    \advance\dimen\charbox by -\dp\slabox
    \advance\dimen\charbox by -\ht\charbox
    \advance\dimen\charbox by \dp\charbox
    \divide\dimen\charbox by 2
    \raise-\dimen\charbox\hbox to \wd\charbox{\hss/\hss}
    \llap{$#1$}
}}

\newc{\onegraph}[4]{%
 \unitlength=1in
 \begin{picture}(3,2.3)
    \put(-0.6,0){\epsfig{file=#1.eps, width=3.7in}}
    \put(0.33,0.48){\epsfig{file=#12.eps, width=2.509in}}
    \put(2.9,1.1){\rotatebox{90}{$#2$}}
    \put(1.3,0.2){\makebox(0,0){$#3$}}
    \put(0.05,1.2){\rotatebox{90}{$#4$}}
  \end{picture}
}

\newc{\ssup}{\tilde{u}}
\newc{\ssdown}{\tilde{d}}
\newc{\ssstrange}{\tilde{s}}
\newc{\sscharm}{\tilde{c}}
\newc{\sstop}{\tilde{t}}
\newc{\ssbottom}{\tilde{b}}
\newc{\sse}{\tilde{e}}
\newc{\ssmu}{\tilde{\mu}}
\newc{\sstau}{\tilde{\tau}}
\newc{\ssnue}{\tilde{\nu}_{e}}
\newc{\ssnumu}{{\tilde{\nu}_{\mu}}}
\newc{\ssnutau}{{\tilde{\nu}_{\tau}}}
\newc{\ssbnue}{\bar{\tilde{\nu}}_{e}}
\newc{\ssbnumu}{\bar{\tilde{\nu}}_{\mu}}
\newc{\ssbnutau}{\bar{\tilde{\nu}}_{\tau}}
\newc{\neut}{{\tilde{\chi}}^0}
\newc{\charge}{{\chi}}
\newc{\glu}{\tilde{g}}
\newc{\Higgs}{H^0}
\newc{\Azero}{A_0}

\newc{\nue}{\nu_e}
\newc{\numu}{\nu_{\mu}}
\newc{\nutau}{\nu_{\tau}}
\newc{\bnue}{\bar{\nu_e}}
\newc{\bnumu}{\bar{\nu_{\mu}}}
\newc{\bnutau}{\bar{\nu_{\tau}}}
\newc{\Br}{\mathrm{Br}}
\newc{\stext}[1]{{\color{red}  #1}}

\begin{document}

\title{All Possible Lightest Supersymmetric Particles in R-Parity Violating 
Minimal Supergravity Models \tnoteref{t1}}
\tnotetext[t1]{Preprint: BONN-TH-2008-14}

\author{Herbi~K.~Dreiner} 
\ead{dreiner@th.physik.uni-bonn.de}

\author{Sebastian~Grab \corref{cor1}}
\ead{sgrab@th.physik.uni-bonn.de}
\cortext[cor1]{Corresponding author}

\address{Bethe Center for Theoretical Physics and Physikalisches
  Institut der Universit\"at Bonn,\\ Nu{\ss}allee 12, 53115 Bonn,
  Germany}

\begin{abstract}
  \noindent We investigate, which lightest supersymmetric particles
  can be obtained via a non-vanishing lepton- or baryon-number
  violating operator at the grand unification scale within the R-parity violating
  minimal supergravity model. We employ the full one-loop renormalization group
  equations. We take into account restrictions from the
  anomalous magnetic moment of the muon and $b \rightarrow s \gamma$,
  as well as collider constraints from LEP and the Tevatron.
  We also consider simple deformations of minimal supergravity models.
\end{abstract}


\maketitle
 
\section{Introduction}
In the minimal supersymmetric Standard Model (MSSM)
\cite{Martin:1997ns}, the lightest supersymmetric particle (LSP) is
stable, guaranteed by the discrete symmetry proton hexality, $\Psix$
\cite{Dreiner:2005rd} or R-parity \cite{Barbier:2004ez}. This also
ensures the stability of the proton. For cosmological reasons the LSP
must then be the lightest neutralino
\cite{Ellis:1983ew,Hebbeker:1999pi} and it has been widely studied as
a very promising cold dark matter candidate
\cite{Blumenthal:1984bp}. However, if we drop P$_6$ then there are
further renormalizable interaction operators in the
superpotential
\cite{Allanach:2003eb}
\begin{eqnarray}
W_{\not \text{P}_6} & = & \eps_{ab}\left[\frac{1}{2} \lam_{ijk} L_i^aL_j^b
\bar{E}_k + \lam'_{ijk}L_i^aQ_j^{bx}\bar{D}_{kx}\right]\notag 
\\&&
+\epsilon_{ab}\kappa^i  L_i^aH_u^b
+\frac{1}{2}\eps_{xyz} \lam''_{ijk}
\bar{U}_i^{\,x} \bar{D}_j^{\,y} \bar{D}_k^{\,z} \, .
\label{RPV_superpot}
\end{eqnarray}
In order to ensure the stability of the proton, we must prohibit
either the first three set of terms which violate
lepton-number, or the last set of terms which violate
baryon-number.\footnote{See also Refs.~\cite{Lee:2007fw,Lee:2007qx}
for a $U(1)'$ solution to the proton decay problem with R-parity
violation.}  These terms violate P$_6$ ($\not\!\text{P}_6$) and thus
the LSP is no longer stable. It is then also not restricted to be the
lightest neutralino and can in principle be any supersymmetric (SUSY)
particle
\cite{Barbier:2004ez}
\begin{equation}
\tilde\chi^0_1,\;\tilde\chi^\pm_1,\;\tilde\ell^\pm_{L/Ri},\;\tilde\tau_1,\;
\tilde\nu_i,\;\tilde q_{L/Rj},\;{\tilde b_1}, \;\tilde t_1,\;\tilde g\,.
\label{LSPs}
\end{equation}
Here we have the lightest neutralino and chargino ($\tilde\chi^0_1,\,
\tilde\chi^\pm_1$), a left-/right-handed charged slepton ($\tilde\ell
^\pm_{L/Ri},\;i=1,2$), a sneutrino ($\tilde\nu_i,\;i=1,2,3$), a
left-/right-handed squark ($\tilde q_{L/Rj},\;j=1,2$), and a gluino ($
\tilde g$).  We have separately listed the lightest stau $\tilde
\tau_1$, sbottom $\tilde b_1$, and stop $\tilde t_1$, as they 
have possibly large Yukawa couplings and left-right mixing and are
thus promising LSP candidates. Potential other dark matter
candidates are the axino \cite{Chun:1999cq}, the gravitino
\cite{Buchmuller:2007ui} and the lightest U-parity particle
\cite{Lee:2007fw,Lee:2007qx,Lee:2008pc}.

In the search for supersymmetry at colliders, it is essential to know
the nature of the LSP, because SUSY particles, if produced, normally
cascade decay down to the LSP within the detector.  The LSP
then decays promptly or with a detached vertex if $\Psix$ is
violated. It is thus a central ingredient of almost all SUSY
signatures.

In Eq.~(\ref{LSPs}), we have a bewildering array of potential LSPs. We
thus need a guiding principle in order to perform a systematic
phenomenological analysis. A well motivated restricted framework for
detailed studies of the MSSM is minimal supergravity (mSUGRA)
\cite{Martin:1997ns}. The 124 free MSSM parameters are reduced to only
five,
\begin{equation}
M_0 \, , M_{1/2} \, , A_0 \, , \tan \beta \, , \text{sgn}(\mu) \, ,
\label{mSUGRA_param}
\end{equation}
where $M_0$ ($M_{1/2}$) is the universal supersymmetry
breaking scalar (gaugino) mass and $A_0$ is the universal
supersymmetry breaking trilinear scalar interaction;
all given at the grand unification (GUT) scale: $M_{\rm GUT}$. $\tanb$
is the ratio of the two vacuum expectation values and $\mu$ is the
Higgs mixing parameter. We obtain the masses of the SUSY particles
(sparticles) by running the renormalization group equations (RGEs) for
the SUSY parameters, from $M_{\rm GUT}$ to the electroweak scale $M_Z$
using \texttt{SOFTSUSY} \cite{rpv_softsusy}. In most of the mSUGRA
parameter space the $\tilde{\chi}_1^0$ or the $\tilde{\tau}_1$, is the
LSP \cite{Allanach:2003eb,Allanach:2006st,Ibanez:1984vq}.

In Ref.~\cite{Allanach:2003eb} the ${\not \Psix}$ effects were taken
into account in the RGEs: giving the ${\not \Psix}$ mSUGRA model.
Here, one additional coupling beyond Eq.~(\ref{mSUGRA_param}) is
assumed:
\begin{equation}
\mathbf{\Lam} \in \{\lam_{ijk},\lam'_{ijk},\lam''_{ijk} \}
\quad \text{at} \; M_{\rm GUT} \, .
\label{lam_couplings} 
\end{equation}
We thus have a simple well-motivated framework, in which we can
systematically investigate the nature of the LSP. It is the purpose of
this letter to determine all possible LSPs in the ${\not \Psix}$
mSUGRA model. We also briefly discuss simple deformations of
mSUGRA. This is very important for SUSY searches at the LHC.

\section{Non-$\tilde{\chi}_1^0$ LSP parameter space of mSUGRA models}
\label{sect_mSUGRA}

If a sparticle directly couples to the operator corresponding
to $\mathbf{\Lam}$, the dominant contributions to the RGE of the
running sparticle mass $\tilde{m}$ are
\cite{Allanach:2003eb,Snu_LSP}:\footnote{For third 
generation sparticles we also need to take into account the
contributions from the Higgs-Yukawa interactions. Their effect is
similar to $\mathbf{\Lambda}$ and $\mathbf{h_\Lam}$ in
Eq.~(\ref{RGE}), see Ref.~\cite{Allanach:2003eb}.}
\begin{eqnarray}
16\pi^2 \frac{d (\tilde{m}^2)}{dt} &=& 
- a_i g_i^2 M_i^2 - b g_1^2 {\cal S} + \mathbf{\Lam}^2 {\cal F}
+ c\, \mathbf{h}^2_\mathbf{\Lam}\; ,
\label{RGE} \\[1mm]
\mathbf{h_\Lam} &\equiv& \mathbf{\Lam} 
\times A_0 \qquad \;\;\text{at} 
\,\,\, M_{\rm GUT}\,. 
\label{hLAMBDA}
\end{eqnarray}
Here $g_i$ ($M_i$), $i=1,2,3$, are the gauge couplings (soft breaking
gaugino masses). $t=\ln Q$ with $Q$ the renormalization scale and
$a_i$, $b$, $c$ are constants of $\mathcal{O}(10^{-1}-10^1)$. ${\cal
S}$ and ${\cal F}$ are linear functions of products of two
softbreaking scalar masses and are given explicitly in
Refs.~\cite{Allanach:2003eb,Snu_LSP}.

\begin{table}[h!]
\begin{center}
\begin{tabular}{cc|cc}
 \hline
 coupling $\mathbf{\Lam}$& LSP & coupling $\mathbf{\Lam}$& LSP\\
 \hline
 $\lam_{132}$ & $\tilde{\mu}_R$  & $\lam_{121}$,$\lam_{131}$,$\lam_{231}$ & $\tilde{e}_R$ \\ 
 $\lam'_{ijk}$ & $\tilde{\nu}_i$  & $\lam''_{212}$ & $\tilde{s}_R/\tilde{d}_R$ \\
 $\lam''_{123},\lam''_{213},\lam''_{223}$ & $\tilde{b}_1$  & $\lam''_{323}$ & $\tilde{t}_1$ \\
 \hline
\end{tabular}
\end{center}
\caption{\label{LSP_candidates} All possible non-$\neut_1$ (and
  non-$\stau_1$) LSP candidates in ${\not \Psix}$ mSUGRA via a
  non-vanishing $\mathbf{\Lam}=\mathcal{O}(10^{-1})$, consistent with
  the experimental bounds, {\it cf.} Refs.~\cite{Barbier:2004ez,Allanach:1999ic}.}
\end{table}

The sum of the first two P$_6$-conserving terms in
Eq.~(\ref{RGE}) is negative and thus \textit{increases} $\tilde{m}$
when running from $M_{\rm GUT}$ to $M_Z$. In contrast, the last two
${\not \Psix}$ terms proportional to $\mathbf{\Lam}^2,$ and
$\mathbf{h} ^2_\mathbf{\Lam}$, are always positive and therefore
\textit{decrease} $\tilde{m}$. We thus expect new LSP candidates
beyond $\tilde{\chi}^0_1$, and $\tilde\tau_1$ if these latter terms
contribute substantially. As we shall see below, this is the
case if $\mathbf{\Lam}=\mathcal{O}(10^{-1})$, {\it i.e.}
$\mathbf{\Lam}= \mathcal{O}(g_i)$ \cite{Snu_LSP}. We can strengthen
the (negative) contribution of $\mathbf{h}^2_\mathbf{\Lam}$, by
choosing a negative $A_0$ with a large magnitude; for moderate
positive $A_0$ there is a cancellation in the RGE evolution of
$\mathbf{h}_\mathbf{\Lam}$ \cite{Snu_LSP}. The other terms are not
significantly affected by $A_0$. Note that we also need $M_{1/2}$
($\tan \beta$) large (small) enough to avoid a $\tilde{\chi}_1^0$
($\tilde{\tau}_1$) LSP.

We now investigate the non-$\tilde{\chi}_1^0$ LSP parameter space of
${\not\!\Psix}$ mSUGRA, \textit{i.e.} with one non-vanishing
${\not\!\Psix}$ coupling $\mathbf{\Lam}=\mathcal{O} (10^{-1})$. 
We discuss the obtainable LSPs for the various
${\not\!\Psix}$-operators. We shall focus here on the $LL\bar E$ and $\bar
U\bar D\bar D$ operators.  In previous work in
Refs.~\cite{Allanach:2003eb,Allanach:2006st,Snu_LSP} it was shown that
$\lam'_{ijk}|_{\rm GUT}=\mathcal{O}(10^{-1})$ can lead to a
$\tilde{\nu}_i$ LSP. In general $LQ\bar D$ operators can only lead to
a $\tilde\nu_i$ LSP beyond a $\tilde{\chi}^0_1$ or a $\tilde\tau_1$, in $\not
\Psix$ mSUGRA. We thus refer to Ref.~\cite{Snu_LSP} for a discussion
of the $\tilde{\nu}_i$ LSP via $\lam'_{ijk}|_{\rm GUT} \not =0$. We
also note that Ref.~\cite{Allanach:2003eb,Jack:2005id} gives the
example of an $\tilde{e}_R$ LSP via $\lam_{231}|_{\rm GUT} =
\mathcal{O}(10^{-1})$.

In searching for LSP candidates, we take into account the constraints
from the decay $b\rightarrow s\gamma$ \cite{Barberio:2008fa}, as well
as the anomalous magnetic moment of the muon \cite{Stockinger:2007pe}.
In the figures, we show (dashed yellow) contour lines corresponding to
the $2\sigma$ window for the BR$(b\rightarrow s\gamma)$
\cite{Barberio:2008fa},
\begin{equation}
2.74 \times 10^{-4} < \text{BR} (b \rightarrow s \gamma) < 4.30 \times 10^{-4} \, ,
\label{bsgamma}
\end{equation} 
and (solid green) contour lines corresponding to the $2\sigma$ window for the SUSY 
contributions to the anomalous magnetic moment of the muon
\cite{Stockinger:2007pe}
\begin{equation}
11.9 \times 10^{-10} < \delta a_\mu^{\rm SUSY} < 47.1 \times 10^{-10} \, .
\label{amu}
\end{equation} 
See also Ref.~\cite{Snu_LSP} and references therein. We employ the LEP
exclusion bound on the light Higgs mass \cite{Barate:2003sz}, but we
reduce it by 3 GeV to $m_h > 111.4$ GeV, to account for numerical
uncertainties of {\tt SOFTSUSY} \cite{Allanach:2006st,Allanach:2003jw}.
We use {\tt microOMEGAs1.3.7} \cite{Belanger:2001fz} to calculate 
BR$(b\rightarrow s\gamma)$ and $\delta a_\mu^{\rm SUSY}$.

\begin{figure}
\begin{center}
      \setlength{\unitlength}{1in} 
\includegraphics{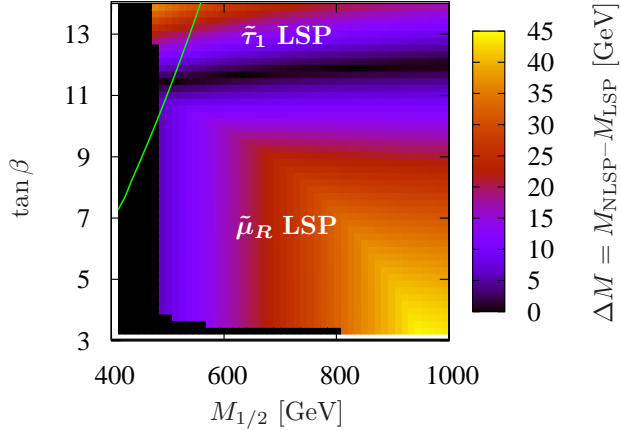}
\caption{\label{lam231_M0M12_delta} Mass
      difference, $\Delta M$, between the NLSP and LSP. The LSP
      candidates are explicitly mentioned in the plot. The
      blackened out region on the left and bottom corresponds to
      parameter points, which posses a tachyon or which violate other
      constraints as described in the text. The contour line is
      described in the text. The other mSUGRA parameter are
      $\lam_{132}|_{\rm GUT}=0.09$, $M_0=170$ GeV, $A_0=-1500$ GeV and
      $\text{sgn}(\mu)=+1$.}
\end{center}
\end{figure}

Our results are summarized in Table~\ref{LSP_candidates} and
are explained in the following.  We only consider operators for which
$\mathbf{\Lam}=\mathcal{O} (10^{-1})$ is consistent with existing
experimental bounds \cite{Barbier:2004ez,Allanach:1999ic}. We
argue that Table~\ref{LSP_candidates} gives a complete list of all
possible non-$\tilde{\chi}_1^0$ and non-$\tilde{\tau}_1$ LSP
candidates in ${\not \Psix}$ mSUGRA. In Sect.~\ref{deform} below, we
shall also consider simple deformations of ${\not \Psix}$ mSUGRA.

\subsection{Non-$\tilde{\chi}_1^0$ LSPs via LLE}
\label{sect_LLE}

The least constrained couplings of the $L_i L_j \bar E_k$ operators,
Eq.~(\ref{RPV_superpot}), are
\cite{Barbier:2004ez,Allanach:2003eb,Allanach:1999ic}
\begin{align}
& \lam_{121},\; \lam_{131} < 0.15 ,\;\;\qquad \qquad\;\; \lam_{123} 
< 0.05 \times (m_{\tilde{\tau}_R}/100\,\mathrm{GeV}) \; , \nonumber \\ 
& \lam_{132},\;\lam_{231} < 0.07 \times 
(m_{\tilde{\mu}_R,\tilde{e}_R}/100\,\mathrm{GeV})\; , 
\label{bounds_LLE}
\end{align}
where the bounds apply at $M_Z$. Note, that $\lam_{ijk}$ is reduced by roughly
a factor of 1.5 when running from $M_{Z}$ to $M_{\rm GUT}$
\cite{Allanach:1999ic}. There is no scaling factor for the first bound as it 
is derived from the neutrino mass bound via the RGEs \cite{Allanach:2003eb}.

We give in Fig.~\ref{lam231_M0M12_delta} the $\tilde{\mu}_R$ LSP
region in the $M_{1/2}$--$\tan \beta$ plane for a
$\lam_{132}$-coupling. We show the mass difference, $\Delta M$,
between the NLSP and LSP. We have employed a lower bound of 190~GeV on
the $\tilde{\mu}_R$ mass to fulfill the strong bound on $\lam_{132}$.
The remaining SUSY particles are then so heavy within ${\not \Psix}$
mSUGRA, that other collider constraints from LEP and the Tevatron are
automatically fulfilled.

\begin{figure}
\begin{center}
      \setlength{\unitlength}{1in} 
      \includegraphics{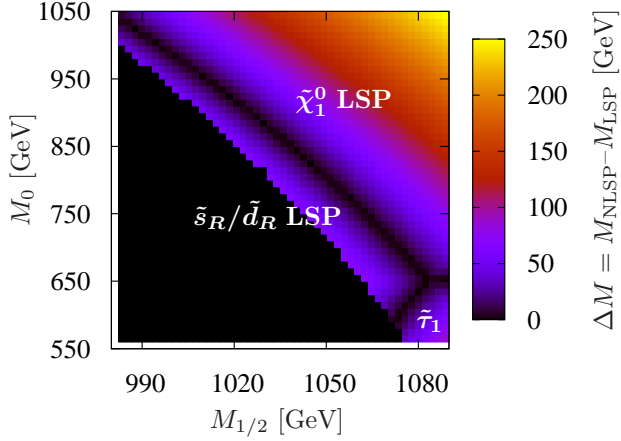}
	\caption{\label{lampp212_M0M12_delta} Same as
	Fig.~\ref{lam231_M0M12_delta}, but with $\lam''_{212}|_{\rm
    GUT}=0.5$, $A_0=-3700$ GeV, $\tanb=19$ and $\text{sgn}(\mu)=+1$.}
\end{center}
\end{figure}

We see that the $\tilde{\mu}_R$ LSP exists in an extended region of
${\not \Psix}$ mSUGRA. We find a $\tilde{\mu}_R$ LSP for all $M_{1/2}
> 480$ GeV, because $M_{1/2}$ increases the mass of the (bino-like)
$\tilde{\chi}_1^0$ faster than the mass of the $\tilde{\mu}_R$
\cite{Snu_LSP}.  The complete $\tilde{\mu}_R$ LSP region in
Fig.~\ref{lam231_M0M12_delta} agrees with BR($b \rightarrow s \gamma$)
at $2\sigma$. But only a tiny region is consistent with $\delta
a_\mu^{\rm SUSY}$ at $2\sigma$, {\it i.e.} lies above the solid green
line.  The mass spectra are rather heavy and thus $\delta a_\mu^{\rm
SUSY}$ is suppressed.

If we use $\lam_{231}$, $\lam_{121}$ or $\lam_{131}$ instead of
$\lam_{132}$ in our parameter scans, we obtain a $\tilde{e}_R$ as the
LSP.  We can not obtain a $\tilde{\ell}_L$ as the LSP in ${\not
\Psix}$ mSUGRA with $\lam|_{\rm GUT} \not =0$.  On the one hand, the
$\Psix$-conserving contributions to the RGEs of $m_{\tilde
{\ell}_L}^2$ have a larger magnitude compared to those for
$m_{\tilde{\ell}_R}^2$.  On the other hand, the (negative) ${\not
\Psix}$ contributions to $m_{\tilde{\ell}_L}^2$ are smaller in
magnitude compared to those for $m_{\tilde{\ell}_R}^2$
\cite{Allanach:2003eb}.

\subsection{Non-$\tilde{\chi}_1^0$ LSPs via UDD}
\label{sect_UDD}

The following baryon-number violating couplings, $\lam''_{ijk}$, are
only constrained by perturbativity
\cite{Barbier:2004ez,Allanach:1999ic,bounds_UDD}
\begin{equation}
\lam''_{212}\,,\lam''_{123}\,,\lam''_{213}\,,\lam''_{223}\,,\lam''_{323} 
\lsim \mathcal{O}(1) \, . 
\label{UDD_couplings}
\end{equation}
The corresponding operators only affect SU(2) singlet squarks directly,
we can thus only obtain $\tilde{q}_R$ LSPs via these $\lam''_{ijk}$
couplings.

We assume that the weak- and mass-eigenstates of \textit{right}-handed
quarks are the same \cite{Agashe:1995qm}. With this assumption we
avoid the RGE generation of additional couplings $\lam''_{lmn}$ at
$M_Z$ out of $\lam''_{ijk}|_{\rm GUT}$, which might be in
contradiction with experiment \cite{Barbier:2004ez,Allanach:1999ic}.
We have also checked that there are then no new ${\not \Psix}$
contributions (at one-loop) in the RGEs which generate off-diagonal
squark mass matrix elements.  The right-handed squark weak-eigenstates
are therefore approximately equal to their mass eigenstates at
$M_Z$. We thus avoid large flavour changing neutral currents.  Note
that we only have experimental information about mixing in the
left-handed quark sector (CKM matrix).

We show in Fig.~\ref{lampp212_M0M12_delta} the $\tilde{d}_R/\tilde{s}
_R$ LSP region via $\lam''_{212}|_{\rm GUT}=0.5$ in the
$M_{1/2}$--$M_{0}$ plane. The $\tilde{d}_R$ and $\tilde{s}_R$ are
degenerate in mass, because both sparticles interact the same via the
gauge interactions and via $\lam''_{212}$ \cite{Allanach:2003eb}.  We
conservatively impose a lower bound of 380 GeV on the 
$\tilde{d}_R/\tilde{s}_R$ mass, consistent with the non-observation of
the $\tilde{d}_R /\tilde{s}_R$ in resonance searches in the dijet
channel at the Tevatron \cite{CDFnote}.\footnote{It is not clear if
Ref.~\cite{CDFnote} can exclude $m_{\tilde{d}_R/\tilde{s}_R} < 380$
GeV. They did not search for single squark resonances. A more detailed
analysis is required, including NLO corrections to single
$\tilde{d}_R/\tilde{s}_R$ production.}
  
\begin{figure}
\begin{center}
      \setlength{\unitlength}{1in} 
      \includegraphics{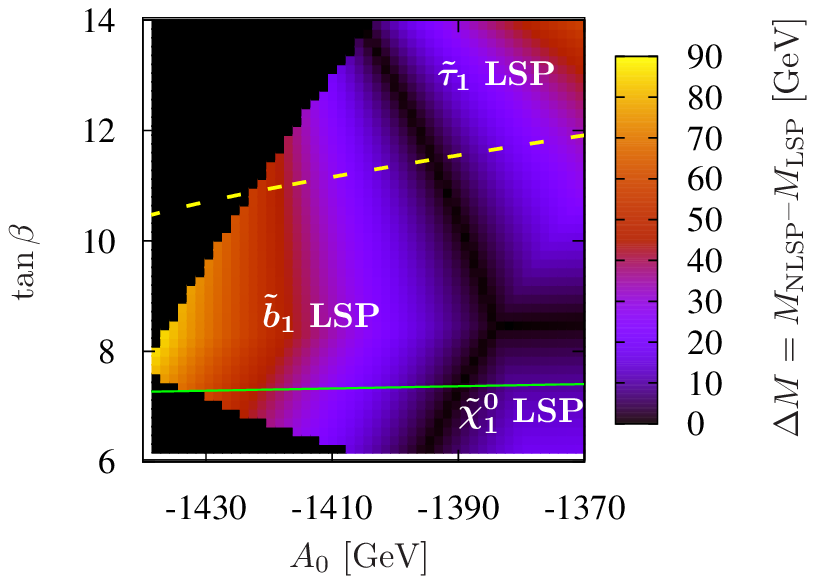}
\caption{\label{lampp223_A0tanb_delta} Same as
  Fig.~\ref{lam231_M0M12_delta}, but with $\lam''_{223}|_{\rm
    GUT}=0.5$, $M_0=120$ GeV, $M_{1/2}=400$ GeV and
  $\text{sgn}(\mu)=+1$.}
\end{center}
\end{figure}

We can not get a $\tilde{c}_R$ LSP via $\lam''_{212}|_{\rm
  GUT}\not=0$.  The ${\not \Psix}$ contributions to the RGEs of the
$\tilde{d}_R$, $\tilde{s}_R$ and $\tilde{c}_R$ mass are the same. But
the $\tilde{c}_R$ couples stronger to the U(1) gaugino than the
$\tilde{d}_R$ and $\tilde{s}_R$ and is therefore always heavier than
$\tilde{d}_R$ and $\tilde{s}_R$.\footnote{The $\tilde{c}_R$ can in principle be lighter than
  the $\tilde{d}_R$ and $\tilde{s}_R$ if $M_{1/2}\lsim 200$ GeV due to
  different D-term contributions.  However, the $\tilde{c}_R$ LSP
  parameter space is in that case excluded by constraints from LEP
  \cite{Heister:2002jc,Barate:2003sz}.} 
For example, the $\tilde{c}_R$ in Fig.~\ref{lampp212_M0M12_delta} is roughly 60 GeV
heavier than the $\tilde{d}_R/\tilde{s}_R$.

Due to $m_{\tilde{d}_R/\tilde{s}_R} > 380$ GeV, we need
$M_{1/2}=\mathcal{O}(1 \, \text{TeV})$, as can be seen in
Fig.~\ref{lampp212_M0M12_delta}, to obtain also a heavy
$\tilde{\chi}_1^0$.  This results in such a heavy mass spectrum that
$\delta a_\mu^{\rm SUSY}$ lies beyond the experimental $2\sigma$
window. However the complete $\tilde{d}_R/\tilde{s}_R$ LSP region in
Fig.~\ref{lampp212_M0M12_delta} is consistent with BR($b \rightarrow s
\gamma$) at $1\sigma$.

Only small $M_{1/2}$ intervals are allowed in
Fig.~\ref{lampp212_M0M12_delta}, because $m_{\tilde{d}_R/\tilde{s}_R}$
at $M_Z$ increases very rapidly with increasing $M_{1/2}$. The
dependence on $M_0$ is weaker, {\it i.e.} $M_0$ intervals up to 100
GeV (for constant $M_{1/2}$) are allowed in
Fig.~\ref{lampp212_M0M12_delta}. These are general features of most of
the squark LSP regions. We thus concentrate on $A_0$ and $\tanb$ in
what follows. $\tanb$ is important, because increasing $\tanb$
increases [decreases] $\delta a_\mu^{\rm SUSY}$ [BR($b\rightarrow
s\gamma$)], \textit{cf.}  Ref.~\cite{Allanach:2006st}.

We give in Fig.~\ref{lampp223_A0tanb_delta} the $\tilde{b}_1$ LSP
region via $\lam''_{223}|_{\rm GUT}=0.5$ in the $A_0$--$\tanb$ plane.
The $\tilde{b}_1$ LSP mass lies between 77 GeV and 180 GeV. The lower
value corresponds to the strongest LEP bound \cite{Heister:2002jc}.
Note, that there is no bound on the $\tilde{b}_1$ LSP mass from
Tevatron searches.  The single $\tilde{b}_1$ production cross section
via $\lam''_{223}|_{\rm GUT}=0.5$ lies below the exclusion limits for
a dijet resonance, {\it cf.} Ref.~\cite{CDFnote}, due to the small
incoming parton luminosity.

Most of the $\tilde{b}_1$ LSP region in
Fig.~\ref{lampp223_A0tanb_delta} is also consistent with BR$(b
\rightarrow s \gamma)$ (below the upper dashed yellow line) and 
$\delta a_\mu^{\rm SUSY}$ (above the lower solid green line) at the $2
\sigma$ level.  We observe that $A_0=\mathcal{O}(-1 \, \text{TeV})$ is
vital to obtain a $\tilde{b}_1$ LSP.  Increasing $A_0$,
\textit{i.e.} decreasing the magnitude, reduces the (negative) effect
of $\lam''_{223}|_{\rm GUT}$ on the running of the $\tilde{b}_1$ mass
and we re-obtain the $\tilde{\chi}_1^0$ or $\tilde{\tau}_1$ LSP,
\textit{cf.}  Ref~\cite{Snu_LSP}.

We can also obtain a $\tilde{b}_1$ LSP, if we use $\lam''_{123}|_{\rm
GUT},\,\lam''_{213}|_{\rm GUT}\not=0$. But now there might be
additional constraints from the Tevatron on di-jet resonances
\cite{CDFnote}.  The couplings $\lam''_{123}$ and $\lam''_{213}$
unlike $\lam''_{223}$ allow for single $\tilde{b}_1$ production via a
valence quark or antiquark, which enhances the hadronic cross section.
Note, that these three couplings can only lead to a $\tilde{b}_1$ LSP,
because the $\tilde{b}_1$ mass (compared to the $\tilde{q}_R$ masses
of the first two generations) is further reduced by the large bottom
Yukawa coupling and by larger left-right mixing.

\begin{figure}
\begin{center}
      \setlength{\unitlength}{1in} 
      \includegraphics{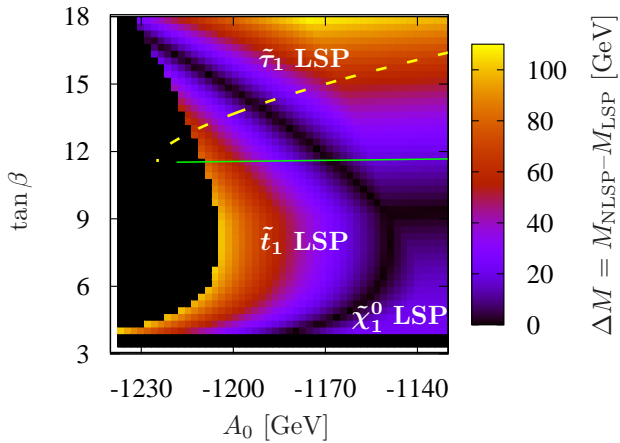}
      \caption{\label{lampp323_A0tanb_delta} Same as
        Fig.~\ref{lam231_M0M12_delta}, but with $\lam''_{323}|_{\rm
          GUT}=0.35$, $M_0=120$ GeV, $M_{1/2}=480$ GeV and
        $\text{sgn}(\mu)=+1$.}
\end{center}
\end{figure}
 
For $\lam''_{323}|_{\rm GUT}=0.35$, we obtain a $\tilde{t}_1$ LSP
as shown in Fig.~\ref{lampp323_A0tanb_delta} for the $A_0$--$\tanb$
plane. The $\tilde{t}_1$ LSP mass ranges from 94 GeV to 200 GeV. The
lower bound corresponds to the LEP bound on $m_{\tilde{t}_1}$
\cite{Heister:2002jc}.\footnote{Unlike the $b_1$ LSP, the $\tilde{t}_1$ LSP has a
  large left-handed component due to left-right mixing. As a
  conservative approach, we take the (stronger) mass bounds from
  Ref.~\cite{Heister:2002jc} for purely left-handed up-type squarks.}
  The $\tilde{t}_1$ LSP region below the upper dashed yellow line
  [above the solid green line] is also consistent with BR($b
\rightarrow s \gamma$) [$\delta a_\mu^{\rm SUSY}$] at $2\sigma$.

We need in general a smaller coupling $\lam''_{ijk}|_{\rm GUT}$ to
obtain a $\tilde{t}_1$ LSP than $\tilde{b}_1$ LSP, because the
$\tilde{t}_1$ mass is further reduced by the large top Yukawa
coupling. This effect is enhanced by a negative $A_0$ with a large
magnitude.  $A_0=\mathcal{O}(-1 \, \text{TeV})$ also leads to large
left-right mixing, which further reduces the $\tilde{t}_1$ mass.  For
the same reasons we can not obtain another squark LSP than the
$\tilde{t}_1$ via $\lam''_{323}|_{\rm GUT}\not=0$.

The complete $\tilde{t}_1$ LSP region in
Fig.~\ref{lampp323_A0tanb_delta} should be testable at the Tevatron
\cite{Choudhury:2005dg}. (See also Ref.~\cite{Dreiner:1991dt}.) The 
authors found that $\tilde{t}_1$ masses up to 190 GeV (210 GeV) can
probably be explored at the Tevatron for an integrated 
luminosity of 2 $\text{fb}^{-1}$ (8 $\text{fb}^{-1}$).  However, this
analysis has not yet been performed by the Tevatron collaborations.
  
\section{LSP candidates in simple deformations of mSUGRA}
\label{deform}

Up to now, we have considered the restricted framework of ${\not
  \Psix}$ mSUGRA. In this section, we want to briefly comment on how
the nature of the LSP can change when we relax some of the mSUGRA
boundary conditions, Eqs.~(\ref{mSUGRA_param}) and
(\ref{lam_couplings}). Throughout this section, we assume that
$\mathbf {\Lambda}$, Eq.~(\ref{lam_couplings}), is
$\lsim\mathcal{O}(10^{-2})$.  ${\not \Psix}$ terms then have no
significant impact on the RGE running of the sparticle masses, {\it
  cf.}  Eq.~(\ref{RGE}). We thus effectively explore the various
corners of deformed, $\Psix$-conserved mSUGRA parameter space. This
has hitherto not been done, since it leads to cosmologically not
viable LSPs in the $\Psix$ case.  The additional effects on the
low-energy mass spectrum of the $\not\!\Psix$-operators for larger
$\mathbf{\Lambda}$ have been discussed in the previous sections, and
apply here correspondingly. In Table~\ref{non_mSUGRA}, we give
examples for the scenarios which we now discuss.

\begin{table*}[t!]
\begin{center}
\begin{tabular}{c|c|c|c}
 \hline
 scenario & universal masses & non-universal masses & LSP \\
 \hline
 1 & $M_0=1500$ GeV, $M_{1/2}=2000$ GeV, & $M_3=300$ GeV & $\tilde{g}$\\[1mm]
 & $A_0=800$ GeV, $\tan\beta=10$, sgn($\mu$)=+1 & \\	
 \hline
 2 & $M_0=180$ GeV, $M_{1/2}=1000$ GeV, & $M_3=400$ GeV & $\tilde{t}_1$ \\[1mm]
 & $A_0=-900$ GeV, $\tan\beta=10$, sgn($\mu$)=+1 & \\
 \hline
 3 & $M_0=100$ GeV, $M_{1/2}=500$ GeV, & $M_{\tilde{E}1}=0$ GeV & $\tilde{e}_R$ \\[1mm]
 & $A_0=0$ GeV, $\tan\beta=10$, sgn($\mu$)=+1 & \\
 \hline
 4 & $M_0=500$ GeV, $M_{1/2}=600$ GeV, & $M_1=1300$ GeV, & $\tilde{\nu}_\tau$ \\[1mm]
 & $A_0=200$ GeV, $\tan\beta=20$, sgn($\mu$)=+1 & $M_{\tilde{L}1}=M_{\tilde{L}2}=M_{\tilde{L}3}=0$ GeV  & \\
 \hline
 5 & $M_0=1000$ GeV, $M_{1/2}=1000$ GeV, & $M_3=300$ GeV, & $\tilde{b}_1$ \\[1mm]
 & $A_0=-800$ GeV, $\tan\beta=40$, sgn($\mu$)=+1 & $M_{\tilde{D}1}=M_{\tilde{D}2}=M_{\tilde{D}3}=0$ GeV & \\
 \hline
 6 & $M_0=1300$ GeV, $M_{1/2}=1100$ GeV, & $M_3=180$ GeV, & $\tilde{d}_R$ \\[1mm]
 & $A_0=400$ GeV, $\tan\beta=30$, sgn($\mu$)=+1 & $M_{\tilde{D}1}=0$ GeV &\\
 \hline
 7 & $M_0=400$ GeV, $M_{1/2}=400$ GeV, & $M_{Hu}^2=1150^2$ GeV$^2$ & $\tilde{t}_1$\\[1mm]
 & $A_0=-1200$ GeV, $\tan \beta=15$, sgn($\mu$)=+1 & & \\
 \hline
\end{tabular}
\end{center}
\caption{\label{non_mSUGRA} Examples for LSP candidates in simple 
  deformations of mSUGRA in 7 different scenarios. The second
  column shows the universal soft breaking masses at $M_{\rm GUT}$. We
  give in the third column the non-mSUGRA soft breaking masses at $M_
  {\rm GUT}$ which deviate from the boundary conditions in the second
  column. In particular, we show the masses of the bino $M_1$, the 
  gluino $M_3$, the right-handed selectron $M_{\tilde{E}1}$, the 
  left-handed sleptons $M_{\tilde{L}i}$, the right-handed down-type 
  squarks $M_{\tilde{D}i}$, and the up-type Higgs $M_{Hu}$. We assume
  the effect of $\mathbf{\Lambda}$, Eq.~(\ref{lam_couplings}), to be 
  negligible, {\it i.e.} $\mathbf{\Lambda} \lsim \mathcal{O}(10^{-2})$ .}
\end{table*}

First, we consider non-universal gaugino masses, {\it i.e.} $M_1 \not
= M_2 \not = M_3$ at $M_{\rm GUT}$. Here, $M_1$, $M_2$, and $M_3$ are
the masses of the bino, the winos, and the gluinos, respectively.  For
$M_3 \ll M_1, M_2$ at $M_{\rm GUT}$, we can obtain a $\tilde{g}$ LSP
if the scalar sparticles and the $\tilde{\chi}_1^0$ are heavy enough,
\textit{i.e.} if $M_0$, $M_1$, and $M_2$ and are large enough. As an
example, we present scenario~1 in Table~\ref{non_mSUGRA}. If we reduce
$M_0$ but maintain $M_3 \ll M_1, M_2$, we get a $\tilde{t}_1$ as the
LSP, see scenario~2.  The squarks are relatively light for small $M_3$
and $M_0$. The $\tilde{t}_1$ mass is further reduced by the effect of
the large top Yukawa coupling on the running and due to left-right
mixing.  For $M_2 \ll M_1$ at $M_{\rm GUT}$, we can get a wino-like
$\tilde{\chi}_1^0$ LSP (instead of a bino-like $\tilde{\chi}_1^0$ as
in mSUGRA) which is nearly degenerate in mass with the
$\tilde{\chi}_1^+$.  A $\tilde{\chi}_1^+$ LSP is in principle possible
but difficult to obtain, see Ref.~\cite{Kribs:2008hq} for details.

Next, we consider non-universal sfermion masses at $M_{\rm GUT}$. For
small right-handed slepton soft breaking masses, $M_{\tilde{E}i}$, we
can obtain a $\tilde{e}_R$ or a $\tilde{\mu}_R$ LSP (beyond the $
\tilde{\tau}_1$ LSP in mSUGRA) if $M_{1/2}$ is large enough,
scenario~3.  For small left-handed slepton soft breaking masses, $M_
{\tilde{L}i}$, and $M_2 \ll M_1$, we can get a $\tilde{\nu} _i$ LSP.
We show the example of a $\tilde{\nu}_\tau$ LSP scenario in
Table~\ref{non_mSUGRA} (scenario~4). A $\tilde{\ell}_{Li}$ LSP is not
possible, because it is always heavier than the $\tilde{\nu}_i$ due to
the different D-terms \cite{Snu_LSP}.  $M_2 \ll M_1$ is vital for the
$\tilde{\nu}_i$ to be the LSP, because $M_{\tilde{L}i}$ increases
faster with $M_2$ then the $\tilde{\chi}_1^0$ mass with $M_1$.
Finally, squark LSPs are possible for small and non-universal squark
soft breaking parameters and $M_3 \ll M_1, M_2$. In these scenarios a
$\tilde{t}_1$ and $\tilde{b}_1$ LSP is preferred, because their masses
are additionally reduced by large Yukawa couplings affecting the RGE
running and by left-right mixing; see scenario~5 in
Table~\ref{non_mSUGRA}.  We obtain non-$\tilde{t}_1$ and
non-$\tilde{b}_1$ squark-LSPs, if we assume non-universal masses for
different squark flavours, {\it cf.}  scenario~6 with a $\tilde{d}_R$
LSP.

Choosing soft breaking Higgs mass parameters different from the
universal scalar mass $M_0$ has the following impact on the sparticle
mass spectrum. On the one hand, the RGE running of third generation
masses is affected due to terms proportional to the Higgs-Yukawa
couplings.  We then obtain, for example, a $\tilde{t}_1$ LSP, {\it
  cf.} scenario~7 in Tab.~\ref{non_mSUGRA}.  On the other hand, the
Higgs mixing parameter $\mu$ and the physical Higgs masses can be
changed. If $\mu$ is small we can get a Higgsino-like
$\tilde{\chi}_1^0$ LSP.  Note that $\mu$ depends on the Higgs soft
breaking masses via radiative electroweak symmetry breaking
\cite{Ibanez:1982fr}.

The mass spectra of the scenarios described above can significantly
change if a large ${\not \Psix}$ coupling is present at $M_{\rm GUT}$,
{\it i.e.}  $\mathbf{\Lambda} = \mathcal{O}(10^{-1})$. The masses are
then modified according to the discussion in the previous sections.
For example, if we assume in scenario~2 in Tab.~\ref{non_mSUGRA} an
additional ${\not \Psix}$ coupling $\lambda_{132}|_{\rm GUT}=0.14$, we
obtain a scenario with a $\tilde{\mu}_R$ LSP, {\it cf.}
Sect.~\ref{sect_LLE}, and a $\tilde{t}_1$ as the NLSP.

\section{Conclusion}
\label{conclusion}

We have investigated for the first time all possible non-$\neut_1$
and non-$\stau_1$ LSPs in 
R-parity violating mSUGRA models; see
Table~\ref{LSP_candidates}.  We found that a non-vanishing $L_i L_j
\bar E_k$ operator at the GUT scale can lead to a $\tilde{e}_R$
($i=1$) or $\tilde{\mu}_R$ ($i=2$) LSP; {\it cf.}
Fig.~\ref{lam231_M0M12_delta}.  A non-vanishing $L_i Q_j \bar D_k$
operator can lead to a $\tilde{\nu}_i$ LSP; {\it cf.}
Ref.~\cite{Snu_LSP}.  We can also obtain squark LSPs, namely the
$\tilde{s}_R$, $\tilde{d}_R$, $\tilde{b}_1$ and $\tilde{t}_1$ via a
non-vanishing $\bar U_i \bar D_j \bar D_k$ operator; see
Fig.~\ref{lampp212_M0M12_delta}, Fig.~\ref{lampp223_A0tanb_delta} and
Fig.~\ref{lampp323_A0tanb_delta}, respectively.
We found $\tilde{\mu}_R$, $\tilde{\nu}_i$, $\tilde{b}_1$ and
$\tilde{t}_1$ LSP scenarios consistent with the observed anomalous
magnetic moment of the muon. All LSP candidates found here can be
consistent with $b \rightarrow s \gamma$ as well as
with collider constraints from LEP and the Tevatron.
According to Ref.~\cite{Choudhury:2005dg}, $\tilde{t}_1$ LSPs up to a
mass of 190 GeV can be tested at the Tevatron with 2 $\text{fb}^{-1}$
of data. We therefore want to encourage the Tevatron collaborations to
investigate the $\tilde{t}_1$ LSP parameter space of R-parity violating 
mSUGRA, as well as to look for squark LSP resonances in dijet events.

We have also discussed simple deformations of mSUGRA;
see Tab.~\ref{non_mSUGRA} for explicit examples. 
We have first assumed that the R-pariy violating coupling at the GUT
scale is small, {\it i.e.} $\mathbf{\Lambda} \lsim \mathcal{O}(10^{-2})$. 
We have found scenarios with a $\tilde{g}$ and $\tilde{t}_1$ LSP if $M_3 \ll M_1, M_2$. 
We can obtain a $\tilde{\ell}_R$ LSP for small right-handed slepton 
(soft breaking) masses. A $\tilde{\nu}_i$ LSP is possible 
for small left-handed slepton masses as long as $M_2 \ll M_1$. 
These scenarios will be significantly affected 
by R-parity violating terms in the RGEs when $\mathbf{ \Lambda} = 
\mathcal{O}(10^{-1})$ as described in Sect.~\ref{sect_mSUGRA}.

Due to the simplicity of the framework, we have in this first study
restricted ourselves mainly to the mSUGRA case. It would be interesting 
to extend this work to other supersymmetry breaking models such as gauge 
mediation \cite{Giudice:1998bp} or anomaly mediation \cite{anomaly_break}.  

\section*{Acknowledgments}

We thank Benjamin Allanach for help with the ${\not \Psix}$ version of
{\tt SOFTSUSY} and Volker B\"uscher for helpful discussions on the
Tevatron searches. SG thanks the `Deutsche Telekom Stiftung' and the
`BCGS of Physics and Astronomy' for financial support. The work of HD
was supported by the SFB TR-33 `The Dark Universe'.


\bibliographystyle{h-physrev}


\end{document}